# Three Patterns to Support Empathy in Computer-Mediated Human Interaction

**Michael J. Lyons**
ATR IRC & MIS Labs
2-2-2 Hikaridai, Keihanna Science City
Kyoto, Japan, 619-0288
mlyons@atr.co.jp

**Daniel Kluender**
ATR MIS Labs
2-2-2 Hikaridai, Keihanna Science City
Kyoto, Japan, 619-0288
dkluender@web.de

**ABSTRACT**
We present three patterns for computer-mediated interaction which we discovered during the design and development of a platform for remote teaching and learning of kanji, the Chinese characters used in written Japanese. Our aim in developing this system was to provide a basis for embodiment in remote interaction, and in particular to support the experience of empathy by both teacher and student. From this study the essential elements are abstracted and suggested as design patterns for other computer-mediated interaction systems.

**INTRODUCTION**
Here we report three patterns which emerged during our implementation of a system for remote learning and teaching of kanji – the Chinese characters used in written Japanese. Our aim was to improve the quality of the online learning experience by providing a technological basis for "shared feeling" or empathy. Our hypothesis was that augmenting the learning platform with a system for communicating felt bodily experience could help both teacher and student in their efforts to understand each others situation, and thereby both ease the learning process and improve the qualitative experience or "feeling". We suggest that these patterns can be applied to other domains of computer-mediated human interaction beyond the teaching/learning context.

**THREE PATTERNS TO SUPPORT EMPATHY**
Here we list the three proposed patterns and summarize the problem, and solution for each one. A more detailed description is given in the individual pattern descriptions which follow.

**E1: Connection to the Body**
*Problem:* "feeling" is a function of subjective bodily experience [8], yet much of computer-mediated interaction suffers from a sense of disembodiment.

*Solution:* Connect bodies and not only minds, by communicating data from physiological sensors.

**E2: Direct and Intuitive Display**
*Problem:* streaming multiple channels of information creates excessive demands for users. Agents increase the sense of alienation by discarding the direct, transparent exchange of non-verbal information upon which the experience of empathy depends [7].

*Solution:* Effectively designed, simple visualization of the relevant information which act as ambient displays [9] do not require constant attention and do not interfere with the main focus of the interaction – the task at hand.

**E3: Reciprocity**
*Problem:* Unresolved inner forces are created in the interaction if the communication channel is one-way. This does not lead to humane, healthy, or emotionally satisfying interactions.

*Solution:* It can be argued that all of these problems stem from a lack of human empathy and more fundamentally from a lack of reciprocity in the interaction. Our proposed solution is to provide a basis for reciprocity, by avoiding one-way information flows.

The reciprocity pattern has been previously described in another context by Schuemmer [6]. We have reason to believe that reciprocity will be one of the most universal design issues in human-computer-human interaction.

**EXAMPLE: REMOTE KANJI LEARNING**
The three patterns arose during the developing of a system for remote learning via the internet. The context of the interaction involves a student who wishes to learn how to write kanji, and a teacher who guides the students learning process. Student and teacher communicate via shared whiteboards. Each has a Wacom tablet and stylus to enter handwritten kanji to the whiteboards (see figures on the title page of the pattern descriptions). Three physiological signs are monitored during the task: blood volume pressure, respiration, and skin conductance. These variables are a measure of the activity of the sympathetic nervous system and reflect level of arousal or stress. We designed simple, direct visualization of relevant signal properties, with minimal signal processing. Previous work has attempted to apply pattern recognition techniques to classify emotional states on the basis of these signals [5]. However these run into difficulty both because of the complexity of the mapping from emotion to physiological sign and because of the importance of contextual information in understanding emotion. Our hypothesis is that humans can deal with contextual information quite well and that it may be possible to learn to use such signals as sources of non-verbal communication, much as we use tone of voice, facial expression and other forms of body language.

We examined this hypothesis in a preliminary experiment which took the form of a structured lesson. Each student was introduced to the basics of writing kanji including the basic component strokes and the order in which they are drawn. Then the teacher taught the student 5 kanji of an increasing level of difficulty. The lesson was followed by a short quiz. The entire lesson took between 30 minutes and 1 hour. Four unpaid volunteers took part in the experiment (3 students and 1 teacher). The 3 students were well motivated to learn as they were active beginning students of Japanese.

One approach for evaluation of such an interaction platform would be to try to measure some quantitative aspect of learning or teaching performance. For example, we could ask whether the platform helps the students to learn faster and/or retain more. But this would tell us little about the human qualities of the interaction, which is our primary interest. Instead, we adopted a subjective approach, asking for first-person accounts from the participants and observers of the experiment.

After the lesson and quiz we interviewed students and teachers separately. They were asked whether they found the visualized physiological signals meaningful or useful for gauging their own emotional status of that of the other, during the task. Because both the task and the information displays were novel to the volunteer students, they generally felt that they were not able to make extensive use of the information during the lesson. One robust observation was that the skin conductance made the students more aware of their emotional status. This seemed to make them more self-conscious and careful during the interaction with the teacher.

The kanji teacher is a member of our research group and so had greater familiarity with the interaction platform and physiological signal displays. The main observation of the teacher was that the skin conductance became useful in pacing the lesson. By the kanji teachers account, excessive activity in the skin conductance was taken to imply that the level or speed of the lesson was too high for the student and needed to be relaxed.

These first-person accounts and third-person observations were consistent with our hypothesis that embodied interaction can engender empathy in online interaction. More extensive studies are needed to fully characterize the influence of the physiology signal displays on the interaction.

**OUTLOOK**

The first author has long been interested in Christopher Alexander's pattern approach to design. The patterns presented here have been formulated according Alexander's style [2] as well as the book by Borchers on interaction design [4]. This is our first attempt at presenting our work in the patterns format and we believe we can learn a great deal from the sheparding process as well as the writer's workshop. In addition, we would be interested in receiving guidance in coding our proposed patterns in PLML.

**ACKNOWLEDGMENTS**

We would like to thank Chi-ho Chan and G. Chaminda de Silva for their helpful suggestions and the volunteers for participating in the evaluation. A talk about this project was presented as a Technical Sketch at SIGGRAPH'03. This work was supported in part by the Telecommunications Advancement Organization of Japan.

# Three Patterns to Support Empathy in Computer-Mediated Human Interaction

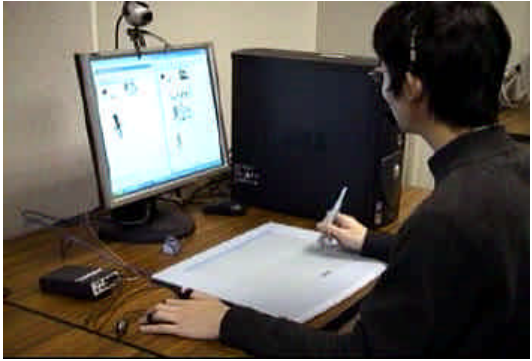 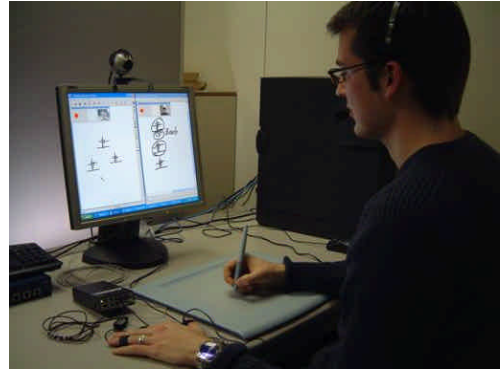

**E1 CONNECTION TO THE BODY**
**E2 DIRECT AND INTUITIVE DISPLAY**
**E3 RECIPROCITY**

*"We can always ask ourselves just how a pattern makes us feel.
And we can always ask the same of someone else."*
*Alexander, The Timeless Way of Building, 1979.*

*"Technologies encalm as they empower our periphery."*
*Weiser & Brown, The Coming Age of Calm Technology, 1996*

# E1 CONNECTION TO THE BODY

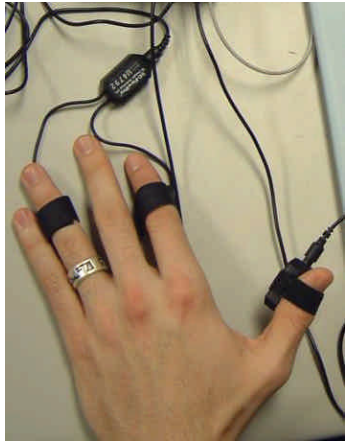 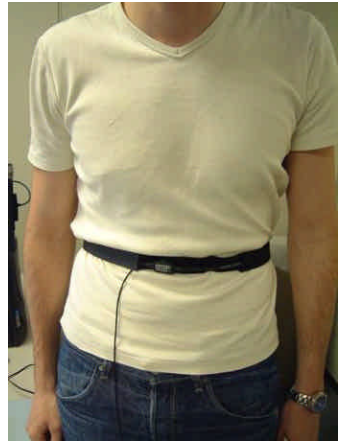

… you are implementing a system for supporting computer-mediated human interaction, for such tasks as remote teaching/learning, cooperative work and/or creativity, or simply connecting with friends in another location. The aim is to increase the sense of the immediacy and reality of another's presence as well as to support greater insight into their state of being.

\*\*\*

**Much of machine-mediated remote interaction suffers from disembodiment – the communication channel is restricted to particular aspects of human communication. By contrast face-to-face communication involves a number of parallel modes of interaction – both verbal and non-verbal. In particular the non-verbal component of communication influences the emotional aspects of our experience of another person.**

Using a telephone, for example, an exchange is restricted to voice, we cannot see or smell or touch the person we are talking with. We adapt to this situation to the point that we are not at all conscious of the fact that this is, in fact, a very restricted form of communication. But this does not mean that it has no effect on the quality of our interaction, or indeed, or experience of the person on the other end of the line, as becomes all too apparent when the topic of conversation carries important emotional content. Increasing the number of modalities through which the users interact improves the situation: for example chat links, can be augmented with video and audio channels.

An additional intriguing, possibility, however, is that the very fact of computer-mediation allows us to create non-verbal communication channels that do not exist even in normal face-to-face communication. For example, biofeedback systems which monitor signals related to stress or arousal, such as pulse, respiration, and skin conductance can be used to create "shared biofeedback", which can allow users to infer each others bodily experience accompanying the interaction.

Therefore:

**Expand the range and quality of information exchanged in a remote interaction, to include non-verbal "body language" not explicitly required in the interaction. Use of physiological signals in particular can allow a group to open windows into each others felt bodily experience associated with the interaction. Our experience of another being as existing and feeling in the world, like ourselves, is a basic component of the experience of the quality of empathy or "shared feeling".**

\*\*\*

An effective visualization of the physiological signals is needed which can be understood without much cognitive load — DIRECT AND INTUITIVE DISPLAY (E2). To engender trust among members of a group one-sided information transfer is to be avoided, this is also a condition for empathetic communication — RECIPROCITY (E3).

# E2 DIRECT AND INTUITIVE DISPLAY

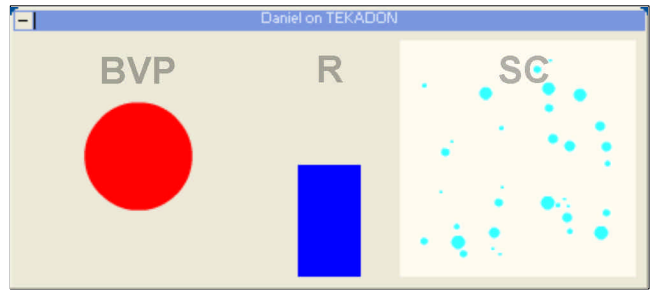

… with a CONNECTION TO THE BODY (E1) there is a need for an effective visualization of the real-time data for each physiological signal.

\*\*\*

**Presenting several channels of real-time physiological data may place unreasonable demands on users' attention, which must anyways be primarily directed at the main focus of the interaction. Furthermore the signals must be presented in an intuitive fashion which is effective even in the periphery or in a quick glance at the display.**

Much of the information in non-verbal communication is processed unconsciously. Facial expressions and body language inform our intuition about a situation even if we have not been consciously focused on these. While we have not evolved mechanisms to process each others vital signs with such ease, an effective visual display can help greatly. In the above example, blood volume pulse, **BVP**, is mapped to the radius of a red circle, so that it resembles a beating heart. Chest expansion, **R**, is mapped to the height of a blue cylinder to represent an expanding and contracting lung. The skin conductance signal, **SC**, was mapped to the density of expanding blue circles to resemble the perspiration of an area of skin.

High-level categorization of "emotional" states is to be avoided in this context mainly for two reasons. The primary reason is that we want signs of felt bodily experience. With a simple, direct visualization of the actual signals, a display of another's vital signs can be identified with ones own felt sensations of pulse, respiration, and perspiration, allowing one insight into how another is feeling. Categorizing the information will interfere with this process. Another reason is simply that emotional states are complex and involve a contextual factor which would be difficult to compute. On the other hand humans are quite capable of taking such contextual factors into account and can become skilled at interpreting emotions from non-verbal cues.

Therefore:

**Create an ambient display which provides simple, intuitive and direct representations of the signals. The depiction must be so transparently clear that these work ambiently, in the periphery and do not demand the user's focused attention. Rudolf Arnheim's work on "Visual Thinking" is a useful resource in the design of such displays.**

\*\*\*

One can learn how to interpret such displays by first associating the visualization in one's display with ones felt experience, and then extending this understanding to the interpretation of others' displays. This process requires that users be able to view their own as well as others' displays – RECIPROCITY (E3).

# E3 RECIPROCITY**

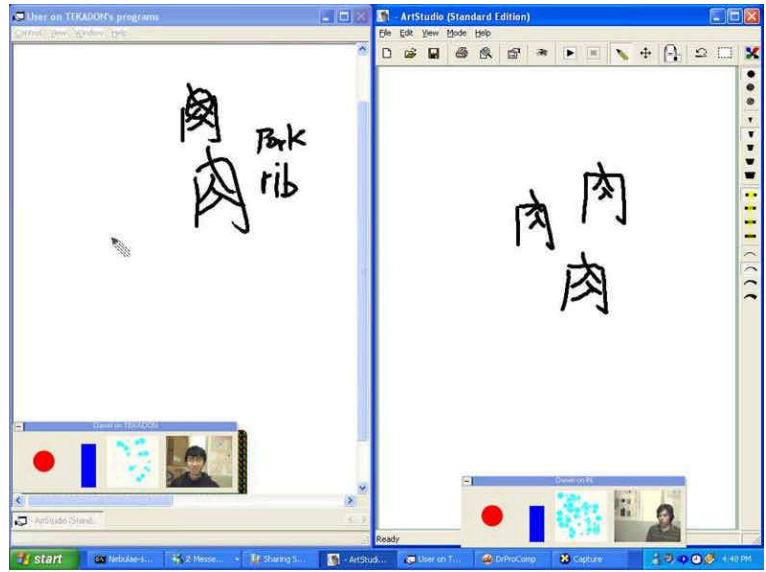

… with signals from an intimate CONNECTION TO THE BODY (E1) being shared in a DIRECT AND INTUITIVE DISPLAY (E2) participants need to establish a relationship of trust.

∗∗∗

**Some people may feel uncomfortable about sharing windows into felt bodily sensations. With one-way interactions, the systems we have described in (E1) and (E2) are reminiscent of the polygraph test used in "lie detection". Who would willingly want to share such information?**

Yet we often share cues to our internal state during face-to-face interaction. There is less of a problem here because, barring such unfortunate situations as one-way mirrors or surveillance cameras, we know who we are interacting with and we have equal access to their non-verbal cues. This reciprocity is basic to establishing trust among participants in such an interaction. This can be encouraged by not allowing a participant to access other participants' vital signs unless they are willing to share their own.

Reciprocity is also a fundamental condition for a full experience of empathy or "shared feeling". I first identify my felt experience with the ambient display of my vital signs. This allows me to read feeling into the ambient display of another's vital signs. The converse is also supported if reciprocity exists. There are further levels to the experience of empathy. Together with the context of the interaction, the ambient display may give me insight into how my actions influence another's felt experience and vice versa. This can continue until I develop a high level of intuition of the other persons felt experience – an experience which some philosophers have termed "intersubjectivity". This condition resolves forces between self and other in computer-mediated interaction bringing the interaction to life.

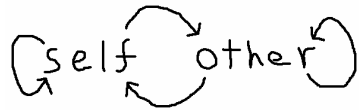

Therefore:

**Actively encourage reciprocity in all interactions. Only if communication is reciprocal can the experience of empathy be felt by both parties, through the feedback loops implicit in the coupling of self and other.**

∗∗∗

This is a basic pattern which has no further references within this pattern language.